\documentstyle [12pt]{article}

\def\K{{\cal K}}

\newcommand{\lbl}[1]{\label{eq: #1}}

\def\R{{\rm I\hspace{-.15em}R}}

\def\b{\begin{equation}} \def\e{\end{equation}}
\def\bd{\begin{displaystyle}} \def\ed{\end{displaystyle}}
\def\ba{\begin{array}} \def\ea{\end{array}}

\def\bee{\begin{enumerate}}
\def\eee{\end{enumerate}}

\def\bes{\begin{eqnarray*}}
\def\ees{\end{eqnarray*}}
\def\be{\begin{eqnarray}}
\def\ee{\end{eqnarray}}

\title{Linear Weyl Gravity in de Sitter Universe}

\author{M.V. Takook$^{1,2}$\thanks{e-mail:
takook@razi.ac.ir}\,\,, M. Reza Tanhayi$^3$\thanks{e-mail:
m$_{-}$tanhayi@iauctb.ac.ir}}
\date{\today}

\begin{document}

\maketitle \centerline{\it $^1$ Department of Physics, Razi
University, Kermanshah, Iran}\centerline{\it $^2$ Groupe de
physique des particules, Universit\'e de Montr\'eal, }\centerline{ \emph{\it C.P.
6128, succ. centre-ville, Montr\'eal, Qu\'ebec,
Canada H3C 3J7}} \centerline{\it $^3$Department of Physics,
Islamic Azad University, Central Tehran Branch, Tehran, Iran}

\begin{abstract}

In this paper we obtain the linear Weyl gravity in de Sitter
background. Its linear form in five-dimensional ambient space notation
has also been obtained. From the group theoretical point of view,
we show that this conformal invariant fourth order theory in its
linear form does not transform as an unitary irreducible
representation of the de Sitter group.

\end{abstract}

\vspace{0.5cm} {\it PACS numbers}: 04.62.+v, 98.80.Cq, 12.10.Dm
\vspace{0.5cm}


\setcounter{equation}{0}
\section{Introduction}

Conformal transformations and conformal techniques have been
widely used in general relativity for a long time (Ref. \cite{1}
and references therein). It has been often claimed that conformal
invariant field theories are renormalizable \cite{f} and conformal
gravity may be an alternative theory of gravity \cite{pm}. \\Since
the gravitational field is long range and seems to travel with the
speed of light, in the linear approximation, at least, its
equations is expected to be conformally invariant. As one knows
Einstein's theory of gravitation is not conformally invariant.
Because of this fact and other issues arising from standard
cosmology and quantum field theory \cite{in}, it seems this theory
could not be considered as a comprehensive universal theory of
gravitational field. Many attempts in generalizing this theory
dating back to the early days of general relativity (for reviewing
see \cite{in1}). The first gravitational theory which is invariant
under the scale transformation was presented by Weyl, hence it is
called Weyl gravity. The Conformal Weyl gravity is based on local
conformal invariance of the metric of the form
$g_{\mu\nu}(x)\rightarrow \Omega^2(x) g_{\mu\nu}(x)$, and leads to
a theory with the field equation of fourth order derivative
(higher-derivative theories). In a series of papers \cite{6},
Mannheim and Kazanas explored the structure of conformal Weyl
gravity as a possible alternative to the Einstein's theory. With
this theory they could potentially explain two outstanding
astrophysical issues namely the cosmological constant \cite{7} and
galactic rotation curve \cite{8} problems. The related works can
be found in Ref.s \cite{9,10,11}. Accordingly, in this work, we
study the cconformally invariant Weyl gravity in de Sitter (dS) space
from the group theoretical point of view.

Gravitational field, in the linear approximation, resembles as a
massless particle with spin-2 that propagates on the background
space-time. According to Wigner's theorem, a linear gravitational
field should transform under the unitary irreducible
representation (UIR) of symmetric group of the
background space-time. In this paper dS space-time has been chosen
as the background. We have shown that linear Weyl gravity can not
be associated with any UIR of dS group. This result may easily
extended for the global scale-invariant action i.e., invariant
under $g_{\mu\nu}(x)\rightarrow\Omega^2 g_{\mu\nu}(x)$ where
$\Omega$ is independent of the space-time coordinates
\cite{wa,de}.

The organization of this paper is as follows. Section $2$ is
devoted to a brief review of the notations. Linear Weyl gravity equation
in the intrinsic dS coordinate is calculated in this section. In
section 3 we project this equation into the five-dimensional
ambient space and we explore the possible relations between this
field and the UIRs of the dS and conformal groups. The results are
summarized and discussed in section 4. Finally, some properties of
the Casimir operators and mathematical relations are presented in
the appendices.

\setcounter{equation}{0}
\section{Notation}

Quantum field theory in dS space has evolved as an exceedingly
important subject, studied by many authors over the course of the
past decade. This is due to the fact that recent astrophysical
data indicate that our universe might currently be in a dS phase.
The importance of dS space has been primarily ignited by the study
of the inflationary model of the universe and quantum gravity. The
de Sitter metric is a solution of the cosmological Einstein's
equation with positive constant $\Lambda$. It is conveniently
described as a hyperboloid embedded in a five-dimensional
Minkowski space
$$
X_H=\{x \in \R^5 ;x^2=\eta_{\alpha\beta} x^\alpha x^\beta
=-H^{-2}=-\frac{3}{\Lambda}\},
$$
where $\eta_{\alpha\beta}=$ diag$(1,-1,-1,-1,-1)$ and $H$ is the
Hubble parameter. The dS metric reads
$$ds^2=\eta_{\alpha\beta}dx^{\alpha}dx^{\beta}=g_{\mu\nu}^{dS}dX^{\mu}dX^{\nu}
,$$ where the $X^\mu$ are  $4$ space-time intrinsic coordinates of
the dS hyperboloid. In this paper we take
$\alpha,\beta,\gamma,\delta,\eta=0,1,2,3,4$ and
$\lambda,\mu,\nu,\rho,\sigma, \tau = 0,1,2,3$. Any geometrical
object in this space can be written in terms of the four local
coordinates $X^\mu$ (intrinsic) or in terms of the five global
coordinates $x^\alpha$ (ambient space).\vspace{.6 cm}

 \textsl{\textbf{I) Fourth order Weyl equation:}}\\
Consider a space-time $({\cal{M}},g_{\mu\nu})$, where $\cal{M}$ is
an $n$-dimensional manifold with metric $g_{\mu\nu}$ of any
signature. The following transformation \b \label{we}
g_{\mu\nu}\rightarrow g'_{\mu\nu}=\Omega^2(x) g_{\mu\nu}, \e where
$\Omega(x)$ is a nonvanishing, regular function, is called a Weyl
or scale transformation and it leaves the light cones unchanged
\cite{wa}. The global scale-invariant action in the metric
signature $(-,+,+,+)$ is \cite{wa} \b I_g=-\frac{1}{4}\int
d^4x\sqrt{-g}\left(a C_{\mu\nu\lambda\rho}C^{\mu\nu\lambda\rho}+b
R^2\right),\e where $a$, $b$ are two dimensionless constant
parameters, $ g=det(g_{\mu\nu})$ and $C_{\mu\nu\lambda\rho}$ is
the conformal Weyl tensor.  The action which is invariant under
the local conformal transformation (\ref{we}) is described by
 $(a=4\alpha_g,b=0)$ \cite{de} \b
I_W=-2\alpha_g\int d^4x\sqrt{-g}
\left(R_{\mu\nu}R^{\mu\nu}-\frac{1}{3}R^2\right)+(\mbox{surface
term}),\e where $\alpha_g$ is a necessarily dimensionless
gravitational coupling constant. Note that in four dimensions
because of Gauss-Bonnet theorem the contribution of
$R_{\mu\nu\lambda\rho}R^{\mu\nu\lambda\rho}$ in to the action can
be written in terms of $R^2$ and $R_{\mu\nu}R^{\mu\nu}$. The total
action is  $I \equiv I_W + I_M$, where $I_M$ is the conformal
matter action (for example see \cite{10}). Functional variation of
the total action with respect to the matter fields yields the
equations of motion while its functional variation with respect to
the metric yields the Weyl field equation. So the Weyl field
equation becomes as follows \b \label{po} W_{\mu\nu}\equiv
W^{(2)}_{\mu\nu}-\frac{1}{3}W^{(1)}_{\mu\nu}=\frac{1}{4\alpha_g}T_{\mu\nu}
,\e where $T_{\mu\nu}\equiv \frac{\delta I_M}{\delta g^{\mu\nu}}$
is the energy-momentum tensor. It is given by Eq. (7) in Ref. \cite{10},
and \footnote{In this paper we follow the sign convention of
Ref. \cite{wa} namely
$$R^a_{bcd}=\partial_c\Gamma^a_{bd}+\Gamma^a_{ec}\Gamma^e_{bd}-c\leftrightarrow
d,$$$$R_{bd}=+R^a_{bad},\,\,\,\mbox{and}\,\,R=+R^a_a.$$}
$$ W^{(1)}_{\mu\nu}=-2g_{\mu\nu}\Box R+2\nabla_\mu\nabla_\nu
R -2R R_{\mu\nu}+\frac{1}{2}g_{\mu\nu}R^2,$$
$$W^{(2)}_{\mu\nu}=-
\frac{1}{2}g_{\mu\nu}\Box R-\Box
R_{\mu\nu}+\nabla_\lambda\nabla_\nu
R_{\mu}^{\lambda}+\nabla_\lambda\nabla_\mu R_{\nu}^{\lambda}-
2R_{\mu}^{\lambda}R_{\nu\lambda}+\frac{1}{2}g_{\mu\nu}R_{\rho\lambda}R^{\rho\lambda}.$$
It can be shown that $W_{\mu\nu}$ is covariantly conserved and
traceless. So this theory leads a fourth order theory of gravity,
it is sometimes known as $R^2$-gravity theory.

\textsl{\textbf{II) Linear field equation in de Sitter space:}}\\
Now we want to obtain the linear approximation of the free field
equation in dS background. In the background field method,
$g_{\mu\nu}=g^{BG}_{\mu\nu}+h_{\mu\nu}$, we suppose
$g^{BG}_{\mu\nu}=g^{dS}_{\mu\nu}$, so we have \b \label{vari}
g_{\mu\nu}=g^{dS}_{\mu\nu}+h_{\mu\nu},\;\;g^{dS}_{\mu\nu}\equiv
\tilde{g}_{\mu\nu},\;\;g^{\mu\nu}=\tilde{g
}^{\mu\nu}-h^{\mu\nu}+{\cal O}(h^2). \e Taking the first variation
of $W_{\mu\nu}$ under the small metric variation (\ref{vari}) and
preserving only linear terms of $h_{\mu\nu}$ results in (appendix
B)

$$  W^{(1)-L}_{\mu\nu}=12H^2\Box h_{\mu\nu}+72H^4 h_{\mu\nu}+2\tilde{g}_{\mu\nu}\Big(\Box^2 h-\Box
\nabla\cdot\nabla\cdot h +3H^2\nabla\cdot\nabla\cdot h-9H^4
h\Big)$$\b+2\Big(\nabla_\mu\nabla_\nu\nabla\cdot\nabla\cdot
h-\nabla_\mu\nabla_\nu\Box h-6H^2\nabla_\lambda\nabla_\mu
h^\lambda_\nu-6H^2\nabla_\lambda\nabla_\nu
h^\lambda_\mu+3H^2\nabla_\mu\nabla_\nu h\big) ,\e

$$
W^{(2)-L}_{\mu\nu}=36H^4h_{\mu\nu}+9H^2\Box
h_{\mu\nu}+\frac{1}{2}\tilde{g}_{\mu\nu}\Big(\Box^2
h-\Box\nabla\cdot\nabla\cdot h-3H^2\Box
h+6H^2\nabla\cdot\nabla\cdot h-18H^4
h\Big)$$$$+6H^2\nabla_\mu\nabla_\nu h+\frac{1}{2}\Box\Big(\Box
h_{\mu\nu}+\nabla_\mu\nabla_\nu h-\nabla_\lambda\nabla_\mu
h^\lambda_\nu-\nabla_\lambda\nabla_\nu
h^\lambda_\mu\Big)-9H^2\Big(\nabla_\lambda\nabla_\mu
h^\lambda_\nu+\nabla_\lambda\nabla_\nu
h^\lambda_\mu\Big)$$$$-\frac{1}{2}\nabla_\lambda\nabla_\mu\Big(\Box
h^\lambda_\nu+\nabla^\lambda\nabla_\mu h-\nabla_\rho\nabla_\mu
h^{\lambda\rho}-\nabla_\rho\nabla^\lambda
h^\rho_\nu\Big)$$\b-\frac{1}{2}\nabla_\lambda\nabla_\nu\Big(\Box
h^\lambda_\mu+\nabla^\lambda\nabla_\nu h-\nabla_\rho\nabla_\nu
h^{\lambda\rho}-\nabla_\rho\nabla^\lambda h^\rho_\mu\Big).\e Thus
the linear free field Weyl equation (\ref{po}) can be written in
de Sitter background as follows \b W_{\mu\nu}^L\equiv\label{wy}
W^{(2)-L}_{\mu\nu}-\frac{1}{3} W^{(1)-L}_{\mu\nu}=0,\e or
$$2W_{\mu\nu}^L=\Box^2h_{\mu\nu}-6H^2\Box
h_{\mu\nu}+8H^4h_{\mu\nu}-H^2\nabla_\mu\nabla\cdot
h_\nu-H^2\nabla_\nu\nabla\cdot
h_\mu+\frac{2}{3}\nabla_\mu\nabla_\nu\nabla\cdot\nabla\cdot
h$$$$+\tilde g_{\mu\nu}\Big(\frac{1}{3}\Box\nabla\cdot\nabla\cdot
h-\frac{1}{3}\Box^2 h+2H^2\nabla\cdot\nabla\cdot h+H^2\Box
h-2H^4h\Big)+\frac{1}{3}\nabla_\mu\nabla_\nu\Box
h$$\b\label{weylds}+2H^2\nabla_\mu\nabla_\nu h
+\frac{1}{3}\nabla_\mu\nabla_\nu\Box h-\nabla_\mu\Box\nabla\cdot
h_\nu-\nabla_\nu\Box\nabla\cdot h_\mu=0.\e where the box operator
$\Box =\tilde{g}_{\mu\nu}\nabla^\mu\nabla^\nu $, is the
Laplace-Beltrami operator in the dS space. In the process of
getting above relation, we used (appendix B) \b \nabla_\lambda
\nabla_\mu h_\nu^\lambda=\nabla_\mu \nabla\cdot
h_\nu+4H^2h_{\mu\nu}-H^2\tilde{g}_{\mu\nu}h,\e \b
\nabla_\lambda\Box h_\rho^\lambda=\Box\nabla\cdot
h_\rho+5H^2\nabla\cdot h_\rho-2H^2\nabla_\rho h,\e \b
\nabla_\lambda\nabla_\nu\Box
h^\lambda_\mu=\nabla_\nu\Box\nabla\cdot
h_\mu+5H^2\nabla_\nu\nabla\cdot h_\mu -2H^2\nabla_\nu\nabla_\mu
h+4H^2\Box h_{\mu\nu}-H^2\tilde{g}_{\mu\nu}\Box h.\e The flat
limit of the Eq. (\ref{weylds}) can be easily obtain by taking the
Hubble constant equal to zero as follows
$$\Box^2h_{\mu\nu}+\frac{2}{3}\partial_\mu\partial_\nu\partial\cdot\partial\cdot
h+\eta_{\mu\nu}\Big(\frac{1}{3}\Box\partial\cdot\partial\cdot
h-\frac{1}{3}\Box^2 h\Big)+\frac{1}{3}\partial_\mu\partial_\nu\Box
h-\partial_\mu\Box\partial\cdot
h_\nu-\partial_\nu\Box\partial\cdot h_\mu=0,$$ in the flat limit
$\Box=\partial_\mu\partial^\mu$ is D'Alembertian operator. Now let
us consider the Eq. (\ref{weylds}), by imposing the traceless and
divergenceless conditions on $h_{\mu\nu}$ i.e.,
$$ h^\mu_\mu= h=0,\;\; \nabla_\mu h^{\mu}_\nu=\nabla \cdot h_\nu=0,$$
which are important in considering the UIR of dS group, we get the
following field equation
 \b \label{weli}\Box^2h_{\mu\nu}-6H^2\Box
h_{\mu\nu}+8H^4h_{\mu\nu}=0.\e In order to explore the possible
relation between this field equation and the UIR of the de Sitter
group in the next section we translate this equation into the
ambient space notation, since in this notation the Casimir
operators of the de Sitter group, SO(1,4), take a simple form and
also the UIRs of the de Sitter group are easy to consider
(appendix A). \setcounter{equation}{0}
\section{Linear field equation in ambient space notation}

In order to clarify the relation between field equation and the
representation of the dS group, we have adopted the tensor field
$\K_{\alpha\beta}(x)$ in ambient space notation. In this notation,
the relationship with UIRs of the dS group becomes straightforward
because the Casimir operators are easily identified with the field
equation \cite{ta}. The transverse tensor field
$\K_{\alpha\beta}(x)$ is locally determined by the ``intrinsic''
field $h_{\mu\nu}(X)$ through \b\lbl{passage}
h_{\mu\nu}(X)=\frac{\partial x^{\alpha}}{\partial
X^{\mu}}\frac{\partial x^{\beta}}{\partial
X^{\nu}}\K_{\alpha\beta}(x(X)). \e The symmetric tensor field
$\K_{\alpha\beta}(x)$ is defined in dS space and will be viewed
here as a homogeneous function, with some arbitrarily chosen
degree $\sigma$, in the $\R^5$-variables $x^{\alpha}$ \cite{di} \b
x^{\alpha}\frac{\partial }{\partial
x^{\alpha}}\K_{\beta\gamma}(x)=x\cdot\partial \K_{\beta\gamma}
(x)=\sigma \K_{\beta\gamma}(x). \e It also satisfies the
transversality condition \cite{di} \b x\cdot\K(x)=0,\mbox{ \it
i.e. }x^\alpha \K_{\alpha\beta}(x)=0,\mbox{ and } x^\beta
\K_{\alpha\beta}(x)=0 . \e  To express tensor field in terms of
the ambient space coordinates, transverse projection is defined \b
(Trpr \K)_{\alpha_1 \cdots \alpha_l}\equiv
\theta_{\alpha_1}^{\beta_1}
\cdots\theta_{\alpha_l}^{\beta_l}\K_{\beta_1 \cdots \beta_l}\;.\e
The transverse projection guarantees the transversality in each
index. Therefore the covariant derivative of a tensor field,
$T_{\alpha_1....\alpha_n}$, in the ambient space notation becomes
\b Trpr \bar\partial_\beta \K_{\alpha_1 ..... \alpha_n}\equiv
\nabla_\beta T_{\alpha_1....\alpha_n}\equiv \bar
\partial_\beta
T_{\alpha_1....\alpha_n}-H^2\sum_{i=1}^{n}x_{\alpha_i}T_{\alpha_1..\alpha_{i-1}\beta\alpha_{i+1}..\alpha_n}.\e
In the above relations $\bar\partial_{\alpha}$ is the tangential
(or transverse) derivative in dS space,$$
\bar\partial_{\alpha}=\theta_{\alpha\beta}\partial^{\beta}=\partial_{\alpha}+H^2x_{\alpha}x\cdot\partial,\,\,\,\,\,x\cdot\bar\partial=0\,,$$
and $\theta_{\alpha\beta}$ is the transverse projector defined by
$\theta_{\alpha\beta}=\eta_{\alpha\beta}+H^2x_{\alpha}x_{\beta}\,$.
Applying this procedure to a transverse second rank tensor field,
leads to \b {\cal T}_{\beta\gamma\eta}\equiv Trpr
\bar\partial_\beta \K_{\gamma\eta}=\bar\partial_\beta
\K_{\gamma\eta}-H^2x_\gamma \K_{\beta\eta}-H^2x_\eta
\K_{\gamma\beta},\e where ${\cal T}_{\beta\gamma\eta}$ is now a
transverse tensor field of rank $3$. For transverse tensor fields
of rank $3,4$ and $5$, we can respectively write \b \label{m}{\cal
M}_{\alpha\beta\gamma\eta}\equiv Trpr \bar\partial_\alpha {\cal
T}_{\beta\gamma\eta}=\bar\partial_\alpha {\cal
T}_{\beta\gamma\eta}-H^2x_\beta {\cal
T}_{\alpha\gamma\eta}-H^2x_\gamma {\cal
T}_{\beta\alpha\eta}-H^2x_\eta {\cal T}_{\beta\gamma\alpha},\e \b
{\cal N}_{\delta\alpha\beta\gamma\eta}\equiv Trpr
\bar\partial_\delta {\cal
M}_{\alpha\beta\gamma\eta}=\bar\partial_\delta {\cal
M}_{\alpha\beta\gamma\eta}-H^2x_\alpha {\cal
M}_{\delta\beta\gamma\eta}-H^2x_\beta {\cal
M}_{\alpha\delta\gamma\eta}-H^2x_\gamma {\cal
M}_{\alpha\beta\delta\eta}-H^2x_\eta {\cal
M}_{\alpha\beta\gamma\delta},\e
$$
{\cal P}_{\epsilon\delta\alpha\beta\gamma\eta}\equiv Trpr
\bar\partial_\epsilon {\cal
N}_{\delta\alpha\beta\gamma\eta}=\bar\partial_\epsilon {\cal
N}_{\delta\alpha\beta\gamma\eta}-H^2x_\delta {\cal
N}_{\epsilon\alpha\beta\gamma\eta}-H^2x_\alpha {\cal
N}_{\delta\epsilon\beta\gamma\eta}-H^2x_\beta {\cal
N}_{\delta\alpha\epsilon\gamma\eta}-H^2x_\gamma {\cal
N}_{\delta\alpha\beta\epsilon\eta}$$ \b\label{p} -H^2x_\eta {\cal
N}_{\delta\alpha\beta\gamma\epsilon}.\e For example replacing
$(4.6)$ in to $(4.7)$ results in
$${\cal M}_{\alpha\beta\gamma\eta}=
Trpr \bar\partial_\alpha Trpr \bar\partial_\beta
\K_{\gamma\eta}=\bar\partial_\alpha\left(\bar\partial_\beta
\K_{\gamma\eta}-2H^2x_{(\gamma} \K_{\eta)\beta}\right)-H^2x_\beta
\left(\bar\partial_\alpha \K_{\gamma\eta}-2H^2x_{(\gamma}
\K_{\eta)\alpha}\right)$$ \b -H^2x_\gamma \left(\bar\partial_\beta
\K_{\alpha\eta}-2H^2x_{(\alpha} \K_{\eta)\beta}\right)-H^2x_{\eta}
\left(\bar\partial_\beta \K_{\alpha\gamma}-2H^2x_{(\gamma}
\K_{\alpha)\beta}\right).\e By using the following relations
\cite{gagata}
$$g^{dS}_{\mu\nu}=\frac{\partial x^\alpha}{\partial
X^\mu}\frac{\partial x^\beta}{\partial X^\nu}
\theta_{\alpha\beta},$$
$$ \nabla_\mu  \cdots \nabla_\rho h_{\lambda_{1} \cdots
\lambda_{l}}=\frac{\partial x^\alpha}{\partial X^\mu} \cdots
\frac{\partial x^\gamma}{\partial X^\rho}\frac{\partial
x^{\eta_1}}{\partial X^{\lambda_1}}\cdots \frac{\partial
x^{\eta_l}}{\partial X^{\lambda_l}} Trpr \bar\partial_\alpha
\cdots Trpr \bar\partial_\gamma \K_{\eta_1 \cdots \eta_l},$$ we
obtain \b \nabla_\mu \nabla \cdot h_\nu=\frac{\partial
x^\alpha}{\partial X^\mu}\frac{\partial x^\eta}{\partial
X^\nu}{\cal M}_\alpha\,^\beta\,_{\beta\eta} , \e \b \label{t} \Box
h_{\mu\nu}=\frac{\partial x^\gamma}{\partial X^\mu}\frac{\partial
x^\eta}{\partial X^\nu}{\cal M}^\alpha\,_{\alpha\gamma\eta},\e \b
\Box^2 h_{\mu\nu}=\frac{\partial x^\gamma}{\partial
X^\mu}\frac{\partial x^\eta}{\partial X^\nu}{\cal
P}^\delta\,_\delta\,^\alpha\,_\alpha\,_{\gamma\eta},\e where
${\cal M}^\alpha\,_{\alpha\gamma\eta}$, ${\cal
M}_\alpha\,^\beta\,_{\beta\eta}$ and ${\cal
P}^\delta\,_\delta\,^\alpha_{\alpha\gamma\eta}$ are easily
calculated from (\ref{m}) and (\ref{p}) by contraction of the
indices as follows
$$ {\cal M}^\alpha\,_{\alpha\gamma\eta}= \Big(\bar
\partial^2 -2H^2\Big)\K_{\gamma\eta}-4H^2x_{(\gamma}\bar\partial\cdot\K_{\eta)},$$
$${\cal M}_\alpha\,^\beta\,_{\beta\eta}= \bar\partial_\alpha\bar\partial\cdot\K_\eta-H^2x_\eta\bar\partial\cdot\K_\alpha,$$
 $$ {\cal P}^\delta\,_\delta\,^\alpha\,_{\alpha\gamma\eta}=\Big(
\bar\partial^4-4H^2\bar\partial^2+4H^4\Big)\K_{\gamma\eta}-4H^2\bar\partial^2x_{(\gamma}\bar\partial\cdot\K_{\eta)}
+32H^4x_{(\gamma}\bar\partial\cdot\K_{\eta)}$$
\b-4H^2x_{(\gamma}\bar\partial^2\bar\partial\cdot\K_{\eta)}+8H^4x_\gamma
x_\eta\bar\partial\cdot\bar\partial\cdot\K,\e where
$2T_{(\alpha\beta)}=T_{\alpha\beta}+T_{\beta\alpha}$. These
formalism enable one to project all equations from intrinsic
coordinates into the ambient space notation. So the linear Weyl
equation in the ambient space notation can be written as follows
$${\cal
P}^\delta\,_{\delta\gamma}\,^\gamma\,_{\alpha\beta}- 6H^2{\cal
M}^\delta\,_{\delta\alpha\beta}+8H^4\K_{\alpha\beta}-H^2{\cal
M}_\alpha\,^\gamma\,_{\gamma\beta}-H^2{\cal
M}_\beta\,^\gamma\,_{\gamma\alpha}+\frac{2}{3} {\cal
P}_{\alpha\beta\gamma\delta}\,^{\gamma\delta}$$\b+\theta_{\alpha\beta}\Big(\frac{1}{3}{\cal
P}_\delta\,^{\delta\gamma\eta}\,_{\gamma\eta}+2H^2 {\cal
M}^{\gamma\eta}\,_{\gamma\eta}\Big)-{\cal
P}_\alpha\,^{\delta}\,_\delta\,^\gamma\,_{\gamma\beta}-{\cal
P}_\beta\,^{\delta}\,_\delta\,^\gamma\,_{\gamma\alpha}=0.\e Note
that the condition $ \K'=\K_\alpha^\alpha=0$ has been imposed.
After some tedious but straightforward algebra, we find that:
$$\Big(\bar\partial^4-10H^2\bar\partial^2+24H^4\Big)\K_{\alpha\beta}
-6H^2x_{(\alpha}\bar\partial^2\bar\partial\cdot\K_{\beta)}
+40H^4x_{(\alpha}\bar\partial\cdot\K_{\beta)}
-8H^2\bar\partial_{(\alpha}\bar\partial\cdot\K_{\beta)}$$$$
+4H^4x_\alpha
x_\beta\bar\partial\cdot\bar\partial\cdot\K+\theta_{\alpha\beta}\Big(
\frac{1}{3}\bar\partial^2\bar\partial\cdot\bar\partial\cdot\K+6H^2\bar\partial\cdot\bar\partial\cdot\K\Big)
+\frac{2}{3}
\bar\partial_\alpha\bar\partial_\beta\bar\partial\cdot\bar\partial\cdot\K$$
\b\label{le}+\frac{4}{3}H^2x_\beta\bar\partial_\alpha\bar\partial\cdot\bar\partial\cdot\K+2H^2
x_\alpha\bar\partial_\beta\bar\partial\cdot\bar\partial\cdot\K-2\bar\partial_{(\alpha}\bar\partial^2\bar\partial\cdot\K_{\beta)}=0.\e
 Useful identity in deriving this equation is
$$x_\alpha\bar \partial^2=\bar
\partial^2x_\alpha-4H^2x_\alpha-2\bar \partial_\alpha.$$ By imposing the
divergencelessness on $\K$, namely $ \bar\partial.\K=0$,
 Eq. (\ref{le}) reduces to \b \label{n} \left(\bar
\partial^4-10H^2\bar \partial^2+24H^4\right)\K_{\alpha\beta}=0.\e
This equation can be written in terms of the second order Casimir
operator of the dS group as \b \label{weyl} \left
(Q_0^2+10Q_0+24\right)\K_{\alpha\beta}=0,\,\,\mbox{or}\,\,\,(Q_0+4)(Q_0+6)\K_{\alpha\beta}=0,\e
where $Q_0 (=-H^{-2}\bar\partial^2)$ is the second order Casimir
operator of the dS group (appendix A). On the other hand it has
been proved that a symmetric tensor field of rank-2 will transform
as an UIR of dS group if it satisfies the following equation
\cite{gagata} \b \label{q0} Q_0\Big(Q_0-2\Big)
\K_{\alpha\beta}=0.\e Clearly the Eq. (\ref{q0}) is not compatible
with that in (\ref{weyl}) of the linear Weyl gravity in dS space.
The direct result is that the linear Weyl gravity in dS space
(Eq. (\ref{weyl})) does not transform as an unitary irreducible representation of
the dS group.

\section{Conclusion and outlook}
Conformal symmetry is indeed one of the most important measures of
assessment of massless field in quantum field theory. If the
graviton does exist and propagates on the light cone due to its
long range effect, it should have zero mass. The light cone
propagation immediately imposes the conformal invariance on the
graviton field equations. In other words gravitational field
should transform under the UIR of conformal group.\\  Although
Weyl gravity is conformally invariant, it was shown that its
linear form in dS space does not lead to any UIR of the dS group.
So Weyl gravitational equation is not suitable to describe quantum
gravitational field in the linear approximation in the dS space.
Therefore, if a highly probable linear conformal quantum gravity
exists, it could not be represented by the Weyl gravity. Other
authors have concluded the same through different methods
\cite{f,fapa}. \\On the other hand Barut and B\"{o}hm \cite{19}
have shown that the value of the conformal Casimir operator for
the UIR must be equal to 9. By using this fact, Bineger et al.
\cite{b} proved that only mixed symmetry tensor field of rank-3
can be a physical representation of the conformal group. In the
previous paper \cite{me2}, we extended this result to dS space
which is in agreement with the conclusion of Fronsdal et al. in
anti-de Sitter space \cite{fh}. In other words if we want the
linear gravitational field equations to be conformally invariant
and the fields to transform according to the UIRs of their
symmetry group as well as conformal group, we should write the
equations with a mixed symmetry rank-3 tensor field which leads to
a theory like $R^3$-gravity theory \cite{m3}.

\vskip 0.5 cm

\noindent {\bf{Acknowledgments}}:  M.V.T. would like to thank
M.B. Paranjape and S. Rouhani for their helpful discussions. We also would like to
thank T. Ramezani, H. Ardahali and E. Ariamand for their interest
in this work. \setcounter{equation}{0}
\begin{appendix}

\section{de Sitter Casimir operators}

In this appendix some properties of the Casimir operators and
their relation with the box operator, $\Box=\tilde
g_{\mu\nu}\nabla^\mu \nabla^\nu $, is presented. We also discuss
the solution of the field equation.
\\ There are two Casimir
operators \b Q^{(1)}=-\frac{1}{2}L^{\alpha\beta}L_{\alpha\beta}=
-\frac{1}{2}(M^{\alpha\beta}+S^{\alpha\beta})(M_{\alpha\beta}+S_{\alpha\beta}),\e
$$
 Q^{(2)}=-W_{\alpha}W^{\alpha}\,,\;\;\;
 W_{\alpha}=-\frac{1}{8}\epsilon_{\alpha\beta\gamma\sigma\eta}L^{\beta\gamma}L^{\sigma\eta},$$
 where
$M_{\alpha\beta}=-i(x_{\alpha}\partial_{\beta}-x_{\beta}\partial_{\alpha})=
-i(x_{\alpha}\bar\partial_{\beta}-x_{\beta}\bar\partial_{\alpha})$
and the symbol $\epsilon_{\alpha\beta\gamma\sigma\eta}$ holds for
the usual antisymmetric tensor. The action of `spin' generator
$S_{\alpha\beta}$ is defined by \cite{ta}
$$S_{\alpha\beta}\K_{\gamma\delta}=-i(\eta_{\alpha\gamma}\K_{\beta\delta}-\eta_{\beta\gamma} \K_{\alpha\delta} +
 \eta_{\alpha\delta}\K_{\beta\gamma}-\eta_{\beta\delta}
 \K_{\alpha\gamma}).$$
Following Dixmier \cite{ms5}, we get a classification scheme using
a pair $(p,q)$ of parameters involved in the following possible
spectral values of the Casimir operators:
\begin{equation}
Q^{(1)}_{p}=\left(-p(p+1)-(q+1)(q-2)\right)I_d ,\qquad\quad
Q^{(2)}_{p}=\left(-p(p+1)q(q-1)\right)I_d\,.
\end{equation}
For simplicity we define $Q^{(1)}_{p}\equiv Q_{p}$. Three types of
UIRs are distinguished for dS group, $SO(1,4)$, according to the
range of values of the parameters $q$ and $p$ \cite{ms5,ms6},
namely: the principal, complementary and discrete series. The flat
limit indicates that in the principal and complementary series the
value of $p$ bears meaning of the spin. For the discrete series
the only representation which has a physically meaningful
Minkowskian counterpart is $p=q$ case. Mathematical details of the
group contraction and the physical principles underlying the
relationship between dS and Poincar\'e groups can be found in Refs
\cite{ms7} and \cite{ms8} respectively. We are interested in
spin-2 representation since in the linear approximation graviton
may be considered as a massless spin-2 field. The spin-$2$ tensor
representations are classified with respect to the UIR of dS group
as follows:
\begin{itemize}
\item[i)] The UIRs $U^{2,\nu}$ in the principal series where
$p=s=2$ and $q=\frac{1}{2 }+i\nu$ correspond to the Casimir
spectral values
\begin{equation}
\langle Q_2^{\nu}\rangle=\nu^2-\frac{15}{4},\;\;\nu \in \R,
\end{equation}
note that $U^{2,\nu}$ and $U^{2,-\nu}$ are equivalent. \item[ ii)]
The UIRs $V^{2,q}$ in the complementary series where $p=s=2$ and
$q-q^2=\mu,$ correspond to
\begin{equation}
\langle  Q_2^{\mu}\rangle=q-q^2-4\equiv
\mu-4,\;\;\;0<\mu<\frac{1}{4}\,.
\end{equation}
\item[iii)] The UIRs $\Pi^{\pm}_{2,q}$ in the discrete series
where $p=s=2$ correspond to
\begin{equation}\label{gr}
\langle Q_2^{q}\rangle=-6-(q+1)(q-2), \;\;q=1, 2.
\end{equation}
\end{itemize}
The ``massless'' spin-2 field in dS space corresponds to the
$\Pi^{\pm}_{2,2}$ and $\Pi^{\pm}_{2,1}$ cases in which the sign
$\pm$, stands for the helicity. In these cases, two
representations $\Pi^{\pm}_{2,2}$, in the discrete series with
$p=q=2$, have a Minkowskian interpretation.

The compact subgroup of conformal group $SO(2,4)$ is $SO(2)\otimes
SO(4)$. Let $C(E;j_1,j_2)$ denote the irreducible projective
representation of the conformal group, where $E$ is the
eigenvalues of the conformal energy generator of $SO(2)$ and
$(j_1,j_2)$ is the $(2j_1+1)(2j_2+1)$ dimensional representation
of $SO(4)=SU(2)\otimes SU(2)$. The representation $\Pi^+_{2,2}$
has a unique extension to a direct sum of two UIRs $C(3;2,0)$ and
$C(-3;2,0)$ of the conformal group, with positive and negative
energies respectively \cite{ms7,19}. The latter restricts to the
massless Poincar\'e UIRs $P^>(0, 2)$ and $P^<(0,2)$ with positive
and negative energies respectively. $ {\cal P}^{ \stackrel{>}
{<}}(0,2)$ (resp. $ {\cal P}^{\stackrel{>}{<}}(0,-2)$)  are the
massless Poincar\'e UIRs with positive and negative energies and
positive (resp. negative) helicity. The following diagrams
illustrate these connections
\begin{equation}
\left.
\begin{array}{ccccccc}
&& {\cal C}(3,2,0)& &{\cal C}(3,2,0)&\hookleftarrow &{\cal P}^{>}(0,2)\\
\Pi^+_{2,2} &\hookrightarrow &\oplus&\stackrel{H=0}{\longrightarrow} & \oplus  & &\oplus\\
&& {\cal C}(-3,2,0)& & {\cal C}(-3,2,0) &\hookleftarrow &{\cal
P}^{<}(0,2),\\
\end{array}
\right.
\end{equation}

\begin{equation}
\left.
\begin{array}{ccccccc}
&& {\cal C}(3,0,2)& &{\cal C}(3,0,2)&\hookleftarrow &{\cal P}^{>}(0,-2)\\
\Pi^-_{2,2} &\hookrightarrow &\oplus&\stackrel{H=0}{\longrightarrow}&\oplus &&\oplus\\
&& {\cal C}(-3,0,2)&& {\cal C}(-3,0,2)&\hookleftarrow &{\cal P}^{<}(0,-2),\\
\end{array}
\right.
\end{equation} where the arrows $\hookrightarrow $ designate unique
extension. It is important to note that the representations
$\Pi^{\pm}_{2,1}$ do not have corresponding flat limit.

The action of the Casimir operators $Q_0 $ and $ Q_2$ on the
symmetric rank-2 tensor field can be written in the more explicit
form \cite{gagata} \b Q_2 \K_{\alpha \beta}=(Q_0-6)\K_{\alpha
\beta}+4x_{(\alpha}\bar\partial\cdot\K_{\beta)}-4\bar\partial_{(\alpha}x\cdot\K_{\beta)}+
2\eta_{\alpha\beta}\K_{\alpha\beta}. \e By imposing the conditions
$\bar\partial\cdot\K=0=\K'$ and $\;x\cdot\K=0$, $\K_{\alpha\beta}$
transforms according to the UIR of the dS group, so we have $$Q_2
\K_{\alpha \beta}=(Q_0-6)\K_{\alpha
\beta}=<Q_2>\K_{\alpha\beta},$$
$$Q_2=<Q_2>I_d\,\,\,\mbox{and}\,\,\,Q_0=(<Q_2>+6)I_d.$$
For the UIR of discrete series we have two possibilities,
$<Q_2>=-6$ and $-4$, which correspond to $\Pi_{2,2}^{\pm}$ and
$\Pi_{2,1}^{\pm}$, respectively; thus for these representations we
have $$Q_0 \K_{\alpha\beta}=0,\;\;\; \Big(Q_0-2\Big)
\K_{\alpha\beta}=0.$$ We consider two cases:
$Q_0\K_{\alpha\beta}\neq 0$ and $(Q_0-2)\K_{\alpha\beta}\neq 0$.
For the first case, we define the rank-2 tensor
$A_{\alpha\beta}\,\, (\equiv Q_0\K_{\alpha\beta})$; it will
satisfy the similar conditions as $\K_{\alpha\beta}$, namely
$\bar\partial\cdot A=0=A'$ and $x\cdot A=0$. Now if
$A_{\alpha\beta}$ obeys the following equation
$$(Q_0-2)A_{\alpha\beta}=0\,\,\,\mbox{or
equivalently}\,\,\,(Q_2+4)A_{\alpha\beta}=0,$$ it will transform
according to the UIR of the dS group $\Pi_{2,1}^{\pm}$. \\And for
the second one, we define the rank-2 tensor $B_{\alpha\beta}\,\,
(\equiv (Q_0-2)\K_{\alpha\beta})$ and similar statements hold for
this case. Similarly, if $B_{\alpha\beta}$ obeys the following
equation
$$ Q_0B_{\alpha\beta}=0\,\,\,\mbox{or
equivalently}\,\,\,(Q_2+6)B_{\alpha\beta}=0,$$ it will transform
according to the UIR of the dS group $\Pi_{2,2}^{\pm}$. Obviously
in either cases $\K_{\alpha\beta}$ obeys the following equation as
well \b \label{kh}
Q_0(Q_0-2)\K_{\alpha\beta}=0,\;\;\;\mbox{or}\;\;(Q_2+4)(Q_2+6)\K_{\alpha\beta}=0,\e
note that $Q_0(Q_0-2)=(Q_0-2)Q_0$. Therefore the fields that
satisfy Eq. (\ref{kh}) transform according to one of the UIR of the
dS group namely, $\Pi_{2,1}^{\pm}$ or $\Pi_{2,2}^{\pm}$. \\The
solutions of the Eq. (\ref{kh}) are important, Since they
correspond to the UIRs of dS group. So the definition an
homomorphism between $\K_{\alpha\beta}$ and a rank-3 mixed
symmetry tensor field $F_{\alpha\beta\gamma}$ (Fierz
representation), leads us to a gravitational field which
transforms according to the UIR of conformal group as well
\cite{m3}. In Ref. \cite{me2,petata} we have considered the possible
solutions for the Eq. (\ref{kh}) by taking $\K_{\alpha\beta}={\cal
D}_{\alpha\beta}\phi$, where ${\cal D}_{\alpha\beta}$ is a
polarization tensor and $\phi$ is a scalar field in dS space.

Now we want to obtain the intrinsic counterpart of the
Eq. (\ref{kh}), in other words we want to project it on the
hyperboloid. From the definition of the Casimir operator in dS
space, one can write $Q_0=-H^{-2}\bar\partial^2$. And also we have
(Eqs. (3.12-13))
$$h_{\mu\nu}=\frac{\partial x^\alpha}{\partial
X^\mu}\frac{\partial x^\beta}{\partial X^\nu}\K_{\alpha\beta},$$
$$\Box
h_{\mu\nu}=H^2\frac{\partial x^\alpha}{\partial
X^\mu}\frac{\partial x^\beta}{\partial
X^\nu}(-Q_0-2)\K_{\alpha\beta},$$
$$ \Box^2
h_{\mu\nu}=H^4\frac{\partial x^\alpha}{\partial
X^\mu}\frac{\partial x^\beta}{\partial
X^\nu}(Q_0+2)(Q_0+2)\K_{\alpha\beta},$$ note that we take
$\bar\partial\cdot\K=0$ in ${\cal
M}^\gamma\,_{\gamma\alpha\beta}$. By using the above relations,
Eq. (\ref{kh}) in the intrinsic coordinate becomes: \b \left(
\Box^2+6H^2\Box+8H^4 \right)h_{\mu\nu}=0,\e it is clear this
equation differs from Eq. (\ref{weli}) of the linear Weyl gravity.

 \setcounter{equation}{0}
\section{Some useful mathematical relations}
The variation of the $W_{\mu\nu}^{(1)}$ and $W_{\mu\nu}^{(2)}$
contains some terms that are difficult to calculate especially in
the curved background. The following expressions are important in
deriving the variation of $W_{\mu\nu}$ \cite{a} \b \delta
\Gamma^a_{bd}=\frac{1}{2}g^{ae}\Big(\nabla_b\delta
g_{ed}+\nabla_d\delta g_{be}-\nabla_e\delta g_{bd}\Big),\e
\b\delta R^a_{bcd}=\frac{1}{2}g^{ae}\Big(\nabla_c\nabla_b \delta
g_{de}+\nabla_c\nabla_d\delta g_{be}-\nabla_c\nabla_e\delta
g_{bd}-\nabla_d\nabla_b\delta g_{ce}-\nabla_d\nabla_c\delta
g_{be}+\nabla_d\nabla_e\delta g_{bc}\Big),\e \b\delta
R_{ab}=\frac{1}{2}g^{cd}\Big(\nabla_c\nabla_a\delta
g_{bd}+\nabla_c\nabla_b\delta g_{ad}-\nabla_c\nabla_d\delta
g_{ab}-\nabla_b\nabla_a\delta  g_{cd}\Big),\e \b\delta
R=g^{ab}g^{cd}\Big(\nabla_c\nabla_a\delta
g_{bd}-\nabla_c\nabla_d\delta g_{ab}\Big)-R^{cd}\delta g_{cd}.\e
Now the calculation of $R^2$ terms in the total action becomes
straightforward
$$\delta(\sqrt{-g}R^2)=2\sqrt{-g}R\delta
R+R^2\delta\sqrt{-g}=$$\b\label{var1} \sqrt{-g}\nabla_c
A^c+\sqrt{-g}\Big(2\nabla^e\nabla^f R-2g^{ef}\Box
R-2RR^{ef}+\frac{1}{3}g^{ef}R^2\Big)\delta g_{ef},\e where
$$A^c\equiv2g^{ab}g^{cd}R(\nabla_a\delta g_{bd}-\nabla_d\delta
g_{ab})-2g^{cb}g^{ad}\nabla_aR\delta
g_{bd}+2g^{ab}g^{dc}\nabla_dR\delta g_{ab}.$$ Similarly we can
write $$ \delta\Big(\sqrt{-g}R^{ab}R_{ab}\Big)=\sqrt{-g}\nabla_c
B^c+$$\b
\label{var2}\sqrt{-g}\Big(\nabla_a\nabla^eR^{af}+\nabla_a\nabla^fR^{ae}-\Box
R^{ef}-g^{ef}\nabla_a\nabla_bR^{ab}-2R_a^fR^{ae}+\frac{1}{2}g^{ef}R_{ab}R^{ab}\Big)\delta
g_{ef},\e where $$B^c=g^{cd}R^{ab}\Big(\nabla_a\delta
g_{bd}+\nabla_b\delta g_{ad}-\nabla_d\delta
g_{ab}\big)-g^{ad}\big(\nabla_aR^{cb}\delta
g_{bd}-\nabla_bR^{cd}\delta
g_{ad}\Big)$$$$-g^{bd}\Big(R^{ac}\nabla_a\delta
g_{bd}+\nabla_bR^{ac}\delta g_{ad}\Big)+g^{dc}\nabla_dR^{ab}\delta
g_{ab}.$$ Note that the first terms in the right hand side of
Eq.s(\ref{var1}-\ref{var2}) have no conurbation in considering of
the equation of motion.\\ The linear form has been obtained by
taking $ \delta g_{ab}=h_{ab},$ and $\delta g^{ab}=-h^{ab}.$ For
example we obtain \b \delta R_{ab}=-\frac{1}{2}(\Box
h_{ab}+\nabla_b\nabla_a h -\nabla_c\nabla_b h^c_a-
\nabla_c\nabla_a h^c_b) ,\e \b \delta R=-\Box h+\nabla^a\nabla^b
h_{ab}-R^{ab}h_{ab},\e

$$2\delta(\nabla_c\nabla_d R_{ab})=$$ $$
\nabla_c\nabla_d(-\Box h_{ab}-\nabla_a\nabla_b h+ \nabla_e\nabla_b
h^e_{a}+ \nabla_e\nabla_a h^e_{b})$$
$$ +R_{be}(\nabla_c\nabla_d h^e_a+\nabla_c\nabla_a
h^e_d-\nabla_c\nabla^e h_{ad})- R_{ae}(\nabla_c\nabla_d
h^e_b+\nabla_c\nabla_b h^e_d-\nabla_c\nabla^e h_{bd})$$
$$- \nabla_c R_{eb}(\nabla_d h^e_a+\nabla_a
h^e_d-\nabla^e h_{ad})- \nabla_c R_{ea}(\nabla_d h^e_b+\nabla_b
h^e_d-\nabla^e h_{bd})$$ $$ -\nabla_d R_{eb}(\nabla_c
h^e_a+\nabla_a h^e_c-\nabla^e h_{ac})- \nabla_d R_{ea}(\nabla_c
h^e_b+\nabla_b h^e_c-\nabla^e h_{bc})$$\b- \nabla_e
R_{ab}(\nabla_c h^e_d+ \nabla_d h^e_c-\nabla^e h_{dc}).\e
Multiplying the above equation by $g^{ca}$ and using the following
relation
$$g^{ca}\delta(\nabla_c\nabla_d R_{ab})=\delta (\nabla_c\nabla_d
R^c_{b})+ h^{ca} \delta(\nabla_c\nabla_d R_{ab}),$$ one obtains the
other form of variations, for example
$$2\delta(\Box
R_{ab})=-2h^{de}\nabla_e\nabla_d R_{ab}-\Box(\Box h_{ab}-
\nabla_b\nabla_a h+ \nabla_c\nabla_b h^c_a+ \nabla_c\nabla_a
h^c_b)$$$$-R_{bc}(\Box h^c_a+\nabla^d\nabla_a h^c_d-
\nabla^d\nabla^c h_{ad})- R_{ac}(\Box h^c_b+\nabla^d\nabla_b
h^c_d- \nabla^d\nabla^c h_{bd}) -2\nabla^d R_{cb}(\nabla_d
h^c_a+\nabla_a h^c_d$$\b-\nabla^c h_{ad})-2\nabla^d
R_{ca}(\nabla_d h^c_b+\nabla_b h^c_d-\nabla^c h_{bd})-2 \nabla_c
R_{ab}\nabla^d h^c_d+ \nabla_c R_{ab}\nabla^c h,\e

$$\delta(\nabla_a\nabla_b R)=\nabla_a\nabla_b(-\Box
h+ \nabla_c\nabla^d h_d^c)-\nabla_a\nabla_b h^{cd} R_{cd}-
\nabla_b h^{cd}\nabla_a R_{cd}$$\b- \nabla_a h^{cd}\nabla_b
R_{cd}-\frac{1}{2}(\nabla_a h^c_b \nabla_c R+ \nabla_b h^c_a
\nabla_c R+\nabla^c h_{ab}\nabla_c R)- h^{cd}\nabla_a\nabla_b
R_{cd}, \e

$$\delta(\Box R)=- h^{ab}
\Box R_{ab}- h^{ab}\nabla_b\nabla_a R$$ \b - \Box^2 h+
\Box\nabla_b\nabla^a h_a^b- \Box h_{ab}R^{ab}-2\nabla^c h^{ab}
\nabla_c R_{ab}- \nabla^c h^b_c\nabla_b R+\frac{1}{2}\nabla^c
h\nabla_c R.\e Now one can easily rewrite these variations in the
dS background noting that de Sitter space is a maximally symmetric
space so Riemannian tensor takes the following form \b
R_{abcd}=H^2(g_{ac}g_{bd}-g_{ad}g_{bc}).\e Ricci tensor and Ricci
scalar become \b
 R_{bd}=3H^2g_{bd},\,\, \mbox{and}\,\,
R=12H^2.\e On the other hand we have
$$[\nabla_c,\nabla_f]h^a_b=R^a_{dcf}h^d_b-R^d_{bcf}h^a_d,$$
this relation in dS background becomes
$$[\nabla_c,\nabla_f]h^a_b=H^2\Big(g^a_c
h_{f b}-g^a_f h_{cb}-g_{bf}h^a_c+g_{bc}h^a_f\Big).$$

\end{appendix}


\begin{thebibliography}{a}
\addcontentsline{toc}{chapter}{Bibliographie}
\bibitem{1} V. Faraoni, E. Gunzing and P. Nardone \emph{" conformal transformations in classical gravitational theories and in
cosmology"}, Fund. Cosmicphys. 20 (1999) 121. arXiv:
gr-qc/9811047.
\bibitem{f} C. Fronsdal, Phys. Rev. D 30 (1984) 2081.
\bibitem{pm} P.D. Mannheim, Astrophys. J. 479 (1997) 659.
\bibitem{in} S. Capozziello, et.al., arXiv: gr-qc:07122980.
\bibitem{in1} B.S. DeWitt, Phys. Rep.,  19C (1975) 295; K.S. Stelle, Gen. Rel. and Grav. 9 No. 4 (1978)
353.
\bibitem{6} P.D. Mannheim and D. Kazanas Phys. Rev. D 44 (1991)417;
P.D. Mannheim, Astrophys. J. 391 (1992) 429; Astrophys. J. 419
(1993)150; Phys. Rev. D. 58 (1998) 103511.
\bibitem{7} P.D. Mannheim, Gen. Relativ. Gravi. 22 (1990) 289.
\bibitem{8} P.D. Mannheim and D. Kazanas, Astrophys. J. 342
(1989) 635.
\bibitem{9} P.D. Mannheim,
{\it Conformal Gravity Challenges String Theory}, arXiv:
0707.2283v1.
\bibitem{10} P.D. Mannheim, Phys. Rev. D. 75 (2007)
124006.
\bibitem{11} P. Chen arXiv: gr-qc 1002.4275; A.
Bhattacharya and et. al., arXiv: gr-qc 0910.1112.
\bibitem{wa} C.W. Misner, K.S. Thorne and J.A. Wheeler \emph{Gravitation}, New York, (1973); R.M. Wald, \emph{General Relativity}, Chicago
University press, (1984).
\bibitem{de} B. S. DeWitt, Relativity, Groups and Topology, C. DeWitt and B. S.
DeWitt, Eds., Gordon and Breach New York (1964).
\bibitem{ta} T. Garidi, J. P. Gazeau, S. Rouhani, and M.V. Takook,
J. Math. Phys. 49 (2008) 032501.
\bibitem{di} P.A.M. Dirac, Annals of Math. 37 (1935-b) 429.
\bibitem{gagata} T. Garidi, J. P. Gazeau and M.V. Takook, J. Math. Phys. 44 (2003)
3838.
\bibitem{fapa} J. Bouchami, M.B. Paranjape, Phys. Rev. D 78 (2008) 044022;
L. Fabbri, M. B. Paranjape, {\it Zero-energy plane waves in
conformal gravity}, arXiv: 0812.2491.
\bibitem{19} A. O. Barut and A. B\"{o}hm, J. Math. Phys. 11 (1970) 2938.
\bibitem{b} B. Binegar, C. Fronsdal, and W. Heidenreich, Phys. Rev. D
27(1983)2249.
\bibitem{me2} M. Dehghani, S. Rouhani, M.V. Takook and M.R. Tanhayi, Phys. Rev. D 77 (2008) 064028.
\bibitem{fh} C. Fronsdal and W.F. Heidenreich, J. Math. Phys. 28 (1987) 215.
\bibitem{ms5} J. Dixmier, Bull. Soc. Math. France 89 (1961) 9.
\bibitem{ms6} B. Takahashi, Bull. Soc. Math. France 91 (1963) 289.
\bibitem{ms7} M. Levy-Nahas, J. Math.
Phys. 8 (1967) 1211.
\bibitem{ms8} H. Bacry, J.M. Levy-Leblond, J. Math.
Phys. 9 (1968) 1605.
\bibitem{m3} M.V. Takook, M.R. Tanhayi and S. Fatemi,
J. Math. Phys. 51 (2010) 032503. arXiv:0903.5249v1.
\bibitem{petata} H. Pejhan, M.R. Tanhayi and M.V. Takook, Int. J. Theor. Phys. 49 (2010) 2263.
\bibitem{a} N.H. Barth and S.M. Christensen, Phys. Rev. D 28 (1983) 1876.

\end{thebibliography}
\end{document}